\documentclass[graybox]{svmult}

\usepackage{mathptmx}       
\usepackage{helvet}         
\usepackage{courier}        
\usepackage{type1cm}        
\usepackage{makeidx}         
\usepackage{graphicx}        
\usepackage{multicol}        
\usepackage[bottom]{footmisc}

\makeindex             
\begin{document}

\title*{Transiting Exoplanets with JWST}
\author{Seager, S., Deming, D., Valenti, J. A.}
\institute{S. Seager \at Dept. of Earth, Atmospheric, and Planetary Sciences, Dept. of Physics, Massachusetts Institute of Technology, 77
Massachusetts Ave., 54-1626, Cambridge, MA, 02139
\email{seager@mit.edu} \and D. Deming  \at NASA/Goddard Space
Flight Center, Planetary Systems Branch, Code 693 Greenbelt, MD 20771
\email{leo.d.deming@nasa.gov} \and J. A. Valenti
\at Space Telescope Science Institute, 3700 San Martin Dr., Baltimore, MD 21218 \email{valenti@stsci.edu}}
%
%
\maketitle
\abstract*{ The era of exoplanet characterization is upon us. For a
subset of exoplanets --- the transiting planets --- physical properties
can be measured, including mass, radius, and atmosphere
characteristics.  Indeed, measuring the atmospheres of a further
subset of transiting planets, the hot Jupiters, is now routine with
the {\it Spitzer Space Telescope}.  The {\it James Webb Space
Telescope} ({\it JWST}) will continue {\it Spitzer's} legacy with its large
mirror size and precise thermal stability.  {\it JWST} is poised for the
significant achievement of identifying habitable planets around bright
M through G stars---rocky planets lacking extensive gas envelopes,
with water vapor and signs of chemical disequilibrium in their
atmospheres. Favorable transiting planet systems, are, however,
anticipated to be rare and their
atmosphere observations will require tens to hundreds of hours of {\it JWST}
time per planet. We review what is known about the physical
characteristics of transiting planets, summarize lessons learned from
{\it Spitzer} high-contrast exoplanet measurements, and give
several examples of potential {\it JWST} observations. }

\abstract{ The era of exoplanet characterization is upon us. For a
subset of exoplanets --- the transiting planets --- physical properties
can be measured, including mass, radius, and atmosphere
characteristics.  Indeed, measuring the atmospheres of a further
subset of transiting planets, the hot Jupiters, is now routine with
the {\it Spitzer Space Telescope}.  The {\it James Webb Space
Telescope} ({\it JWST}) will continue {\it Spitzer's} legacy with its large
mirror size and precise thermal stability.  {\it JWST} is poised for the
significant achievement of identifying habitable planets around bright
M through G stars---rocky planets lacking extensive gas envelopes,
with water vapor and signs of chemical disequilibrium in their
atmospheres. Favorable transiting planet systems, are, however,
anticipated to be rare and their
atmosphere observations will require tens to hundreds of hours of {\it JWST}
time per planet. We review what is known about the physical
characteristics of transiting planets, summarize lessons learned from
{\it Spitzer} high-contrast exoplanet measurements, and give
several examples of potential {\it JWST} observations. }

\section{Introduction}
\label{sec:intro}
\begin{figure}[ht]
\centering
\includegraphics[scale=.45]{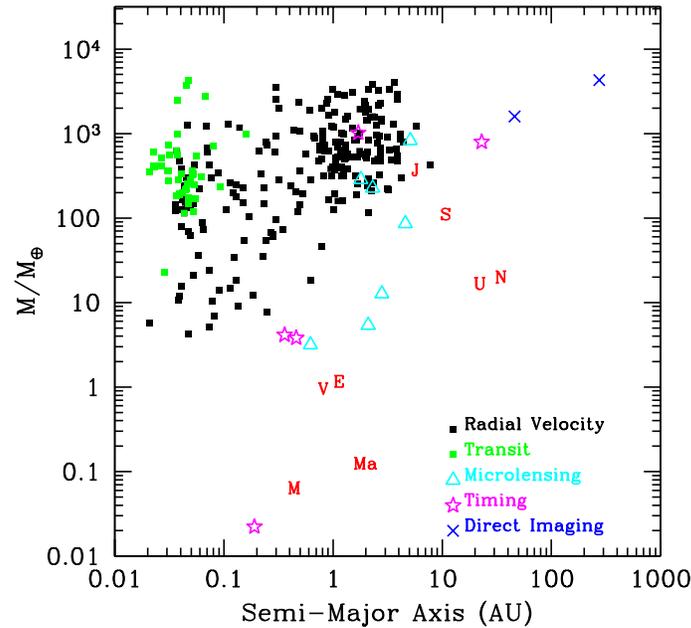}
\caption{Known planets as of July 2008. We have defined planet 
to have a maximum mass of 13 Jupiter masses. The symbols
indicate the discovery technique; see text
for details. Data from \cite{exoe}.}
\label{fig:exocensus}      
\end{figure}

The existence of exoplanets is firmly established with over 300 known
to orbit nearby, sun-like stars. Figure~\ref{fig:exocensus} shows the
known exoplanets as of July 2008 with symbols indicating their
discovery techniques \cite{exoe}.  The majority of the known
exoplanets have been discovered by the Doppler technique which
measures the star's line-of-sight motion as the star orbits the
planet-star common center of mass \cite{butl2006, udry2007}. While
most planets discovered with the Doppler technique are giant planets,
the new frontier is discovery of super Earths (loosely defined as
planets with masses between 1 and 10~$M_{\oplus}$).  About a dozen
radial velocity planets with $M$$<$10~$M_{\oplus}$ and another dozen with 
10~$M_{\oplus}$$<$$M$$<$30~$M_{\oplus}$ have been reported. The transit
technique finds planets by monitoring thousands of stars, looking for
the small drop in brightness of the parent star that is caused by a
planet crossing in front of the star. At the time of writing this
article, around 50 transiting planets are known. Due to selection
effects, transiting planets found from ground-based searches are
limited to small semi-major axes \cite{char2007a}. Gravitational
microlensing has recently emerged as a powerful planet-finding
technique, discovering 6 planets, two belonging to a scaled down
version of our own solar system \cite{gaud2008}. Direct imaging is
able to find young or massive planets with very large semi-major
axes. The mass of directly imaged planets (e.g.,  \cite{chau2004} and
references therein) is inferred from the measured flux based on
evolution models, and is hence uncertain. The timing discovery method 
includes both pulsar planets \cite{wols1992} and planets orbiting stars
with stable oscillation periods \cite{silv2007}.

Many fascinating properties of exoplanets have been uncovered by the
initial data set of hundreds of exoplanets.  A glance at
Figure~\ref{fig:exocensus} shows one of the most surprising features:
that exoplanets exist in an almost continuous range of mass and
semi-major axis.  Not shown in Figure~\ref{fig:exocensus} are the
equally wide range of eccentricities; several different theories for
the origin of planet eccentricities have been proposed.  Because there
is not enough solid material close to the star in a 
protoplanetary disk, the giant planets are believed to have formed
further out in the disk and migrated inwards. The migration stopping
mechanisms, and even the details of planet migration are not fully
understood.  Out of the $\sim$50 known transiting exoplanets, several have
very large radii, and are too big for their mass and age according to
planet evolution models (see Figure~\ref{fig:massradius}). These
``puffed-up'' planets must have extra energy in their core that
prevents cooling and contraction, but no satisfactory explanation
yet exists.

The next step beyond discovery is to characterize the physical
properties of exoplanets by measuring densities, atmospheric
composition, and atmospheric temperatures.

There are two paths to exoplanet characterization. The first is direct
imaging where the planet and star are spatially separated on the
sky. Direct imaging has been successful for discovering hot or massive
planetary candidates with large ($\sim$ 50--100 AU) orbital
and projected spatial separation
\cite{chau2004, chau2005, neuh2005}. Although {\it JWST} will incorporate several coronagraphic
modes, neither the telescope nor the instruments were optimized for
coronagraphy. A relatively large inner working angle and limited
planet-star contrast restrict {\it JWST} to studying young or massive Jupiters
with large semi-major axes.  The case for {\it JWST} coronagraphic observations
of exoplanets is presented in \cite{gard2006}.

Solar-system aged small planets are not observable via direct imaging
with current technology, even though an Earth at 10 pc is brighter
than the faintest galaxies observed by the {\it Hubble Space
Telescope} ({\it HST}). The major impediment to direct observations is
instead the adjacent host star; the Sun is 10 million to 10 billion
times brighter than Earth (for mid-infrared and visible wavelengths,
respectively).  No existing or planned telescope is capable of
achieving this contrast ratio at 1~AU separations.

The second path to exoplanet characterization is via the transit
technique. A subset of exoplanets cross in front of their stars as
seen from Earth (``primary eclipse'' or ``transit''). Planets that
cross in front of their star, also pass behind the star (secondary
eclipse), provided that the transiting planet is on a circular orbit.
The probability to transit is $\sim R_*/a$, where $R_*$ is the stellar
radius and $a$ the semi-major axis. Transits are therefore most easily
found for planets orbiting close to the star. Indeed, all but one of
the $\sim$50 known transiting exoplanets have $a < 0.09$~AU (See
\cite{exoe} and references therein).

\begin{figure}[ht]
\centering
\includegraphics[scale=.5]{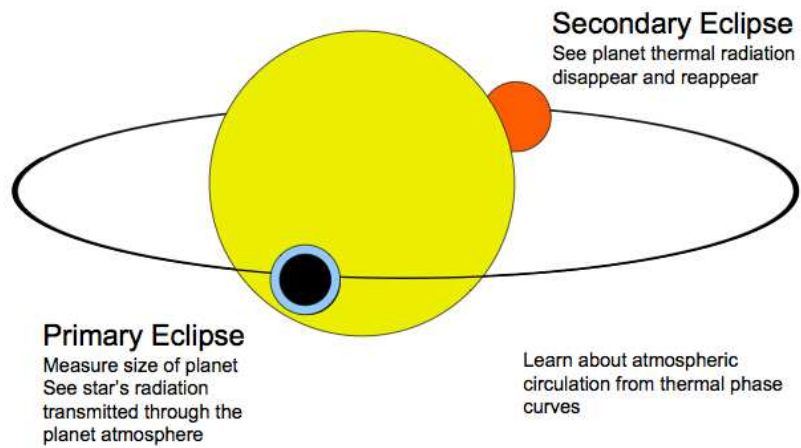}
\caption{Schematic of a planet transit. Not to scale.}
\label{fig:transitchar} 
\end{figure}

Observations of transiting planets exploit separation of photons in
time, rather than in space (see Figure~\ref{fig:transitchar}).  That
is, observations are made in the combined light of the planet-star
system.  When the planet transits the star as seen from Earth the
starlight gets dimmer by the planet-to-star area ratio. If the size of
the star is known, the planet size can be derived.  During the planet
transit, some of the starlight passes through the optically thin part
of the planet atmosphere, picking up spectral features from the
planet. A planetary transmission spectrum can be obtained by dividing
the spectrum of the star and planet during transit by the spectrum of
the star alone (the latter taken before or after transit).

When the planet disappears behind the star, the total flux from the
planet-star system drops. The drop is related to both relative sizes
of the planet and star and their relative brightnesses (at a given
wavelength). The flux spectrum of the planet can be derived by
subtracting the flux spectrum of the star alone (during secondary
eclipse) from the flux spectrum of both the star and planet (just
before and after secondary eclipse). The planet's flux gives
information on the planet composition and temperature gradient (at
infrared wavelengths) or albedo (at visible wavelengths).  Finally,
non-transiting exoplanets can in principle also be observed in the
combined planet-star flux as the planet goes through illumination
phases as seen from Earth. In the thermal infrared the phase
observations provide information on energy redistribution of absorbed
stellar radiation (see Section~\ref{sec-atmsummary}) and at visible
wavelengths the phase observations give information on scattering
particles (gas or clouds).

Primary and secondary eclipses enable high-contrast measurements
because the precise on/off nature of the transit and secondary eclipse
events provide an intrinsic calibration reference.  This is one reason
why the {\it Hubble Space Telescope} and the {\it Spitzer Space
Telescope} have been so successful in measuring high-contrast transit
signals that were not considered in their designs.

\section{{\it Spitzer}'s Legacy}
\label{sec:spitzer}

\subsection{Background}
The {\it Spitzer} Space Telescope is a cryogenically cooled 85 cm diameter
telescope launched into an Earth-trailing orbit in 2003. All three of
{\it Spitzer}'s science instruments (IRAC \cite{irac}, IRS \cite{irs}, and
MIPS \cite{mips}) have been used to study exoplanets. {\it Spitzer} has
revolutionized the field of exoplanets by making measurements of hot
Jupiter atmospheres routine.  The {\it Spitzer} exoplanet studies are
directly relevant to {\it JWST} because {\it JWST} will make similar measurements,
but with much higher S/N, for much smaller exoplanets, or for planets
with semi-major axes beyond 0.05 AU.

\begin{figure}[ht]
\centering
\includegraphics[scale=.5]{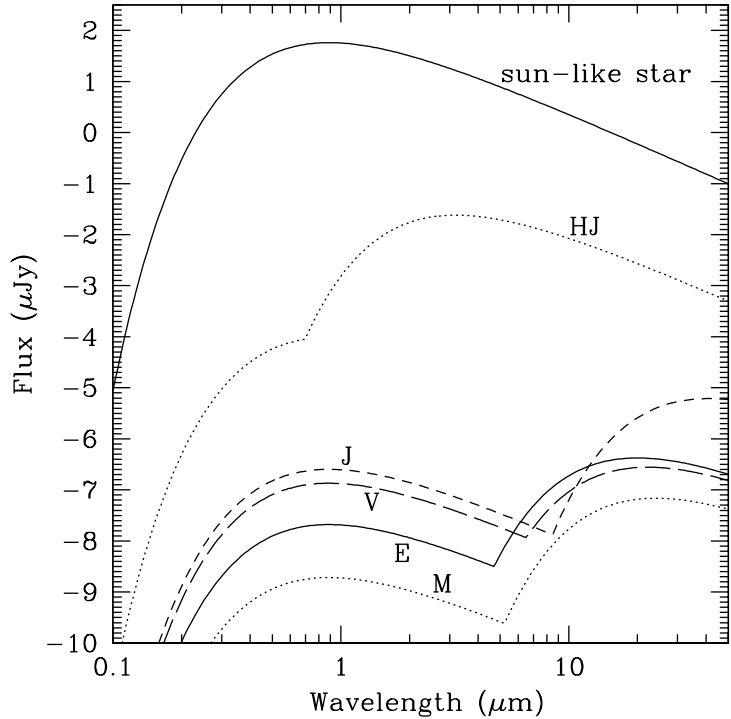}
\caption{Black body flux (in units of 10$^{-26}$ W m$^{-2}$ Hz$^{-1}$)
of some solar system bodies as ``seen'' from 10 pc. The Sun is represented
by a 5750~K black body. The planets Jupiter, Venus, Earth, and Mars
are shown and are labeled with their first initial.  A putative hot
Jupiter is labeled with ``HJ''.  The planets have two peaks in their
spectra. The short-wavelength peak is due to sunlight scattered from
the planet atmosphere and is computed using the planet's
geometric albedo. The long-wavelength peak is from the
planet's thermal emission and is estimated by a black body of the
planet's effective temperature. Data from \cite{cox2000}.
The Hot Jupiter albedo was assumed to be 0.05 and the equilibrium
temperature to be 1600~K.}
\label{fig:blackbody} 
\end{figure}

We describe one reason why {\it Spitzer} has been so successful in
detecting photons from exoplanets during secondary eclipse.
Figure~\ref{fig:blackbody} shows the relative fluxes for the Sun,
Jupiter, Earth, Venus, and Mars, approximating each as a black body.
The planets also reflect light from the Sun at visible wavelengths,
giving them two flux peaks in their schematic spectrum. We see from
Figure~\ref{fig:blackbody} that at infrared wavelengths ( $< 10\mu$m)
the solar system planets are more than 7 orders of magnitude fainter
than the Sun. A generic hot Jupiter, with assumed geometric albedo of
0.05, equilibrium temperature of 1600~K, and a radius of 1.2 $R_J$ is
also shown on the same figure. This representative hot Jupiter is less
than 3 orders of magnitude fainter than the Sun at some wavelengths.
Equally important is that the planet-to-star flux ratio is favorable
where the star and planet flux are high, i.e. plenty of photons are
available to reach the telescope. The $\sim$8~$\mu$m region is
therefore a sweet spot for {\it Spitzer} observations of hot Jupiter
exoplanets.

\subsection{Exoplanet Radii}

A precise planet radius together with planet mass enables a study of
the planet's density and interior bulk composition.  We have shown
that infrared wavelengths are ideal for deriving a precise planet
radius from the transit light curve, due to the miniscule amount
of stellar limb
darkening at infrared wavelengths \cite{rich2006}. In contrast, at
visible wavelengths limb darkening affects the shape of the transit
light curve.  Because limb darkening is imperfectly known for stars
other than the Sun, limb darkening must be solved for from planet
transit light curves at visible wavelengths in order to fit the
planet's radius.  {\it Spitzer} measurements of the HD~209458 primary
eclipse with MIPS at 24~$\mu$m yielded a planetary radius of 1.26$\pm
0.08~R_J$ \cite{rich2006}. At shorter infrared wavelengths with more
photons from the star and where limb darkening is still negligible,
and for host stars of later spectral type than HD~209458, transit
light curves will enable even more precise planet radii to be derived.

Transit observations at infrared wavelengths are especially useful for
planets transiting M stars. M stars are faint at visible wavelengths
with peak flux output at near-IR wavelengths. {\it Spitzer} is arguably the
best existing telescope for determining precise radii of planets
transiting M stars. The Neptune-mass planet GJ~436b \cite{butl2004,
gill2007} was observed by {\it Spitzer} to have a Neptune-like radius of
4.33$\pm 0.18 R_{\oplus}$ \cite{demi2007, gill2007b}.

\subsection{Exoplanet Atmosphere Summary}
\label{sec-atmsummary}
Several different exoplanets have published {\it Spitzer} secondary eclipse
atmosphere measurements. These secondary eclipse measurements have
detection significances ranging from 60$\sigma$ \cite{knut2007} down to
5$\sigma$. Rather than describe each planet individually, we present a
summary based on an important question related to atmospheric
circulation.

Hot Jupiters are expected to be tidally locked to their parent
stars---presenting the same face to the star at all times.  This
causes a permanent day and night side. A long standing question has
been about the temperature difference from the day to night side. Are
the hot Jupiters scorchingly hot on the day side and exceedingly cold
on the night side? Or, does atmospheric circulation efficiently
redistribute the absorbed stellar radiation from the day side to the
night side?

Surprisingly, {\it Spitzer} has found that both scenarios are
possible. {\it Spitzer} has measured the flux of the planet and star system
as a function of orbital phase for several hot Jupiter systems
\cite{harr2006, cowa2007, knut2007}. Assuming that the star is
constant in flux, the resulting brightness change is due to the planet
alone.  The HD~189733 star and planet shows some variation at 8~$\mu$m
during the 30 hour continuous observation of half an orbital phase
\cite{knut2007}. This variation corresponds to about 20\% variation
in planet temperature (from a brightness temperature of 1212 to
973~K).  In contrast, the non-transiting exoplanet Ups And shows a
marked day-night contrast suggesting that the day and night side
temperatures differ by over 1000~K \cite{harr2006}.

Once the stellar radiation is absorbed on the planet's day side, there
is a competition between reradiation and advection. If the radiation
is absorbed high in the atmosphere, the reradiation timescale is short
and reradiation dominates over advection.
In this case the absorbed stellar radiation is reradiated
before it has a chance to be advected around the planet, resulting in
a very hot planet day side and a correspondingly very cold night
side. If the radiation penetrates deep into the planet atmosphere
where it is finally absorbed, the advective timescale dominates and
the absorbed stellar radiation is efficiently circulated around the
planet. This case would generate a planet with a small temperature
variation around the planet. See also \cite{seag2005, harr2007, burr2008, fort2008}. See \cite{show2007} and references therein for a
review discussion of atmospheric circulation models.

\begin{figure}[ht]
\centering
\includegraphics[scale=.55]{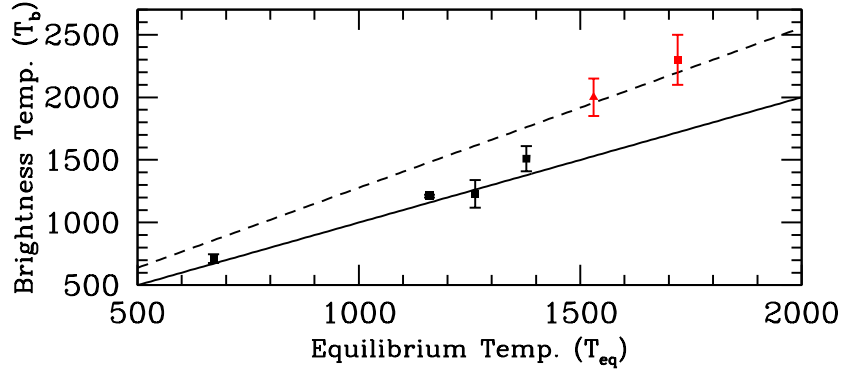}
\caption{ Brightness temperature (8~$\mu$m) as a function of the day
side equilibrium temperature for six hot Jupiters. Brightness
temperature is the measured flux converted to a temperature. The day
side equilibrium temperature $T_{eq}$, is defined in
equation~(\ref{eq:Teq}) and the accompanying text.  The parameter $f$
is used to approximate atmospheric circulation: in this formulation of
equation~(\ref{eq:Teq}), $f=1/4$ corresponds to a uniform temperature
around the planet (solid line), whereas $f=2/3$ corresponds to
instantaneous reradiation on the planet's dayside hemisphere (dashed
line).  The cooler planets lie near the uniform temperature line
whereas the hotter planets lie near the hot day side line.  From left
to right, the brightness temperatures are GJ 436b \cite{demi2007}, HD
189733 b \cite{knut2007}, TrES-1 \cite{char2005}, HD 209458b
\cite{knut2008}, Ups And \cite{harr2006}, HD 149026
\cite{harr2007}. (Note that the Ups And day side temperature is
estimated from its thermal phase curve).}
\label{fig:twoatm} 
\end{figure}

In Figure~\ref{fig:twoatm} we plot measured brightness temperatures of
seven hot exoplanets together with two equilibrium temperature 
($T_{eq}$) curves. One of the $T_{eq}$ curves is for a planet with evenly
redistributed absorbed stellar radiation ($f=1/4$ below),
corresponding to a planet with little temperature difference between
the day and night sides.  The other $T_{eq}$ curve is for a planet
with instantaneous reradiation of absorbed stellar radiation ($f=2/3$
below), corresponding to a planet with a strong day-night temperature
difference. The cooler of the hot exoplanets in
Figure~\ref{fig:twoatm} lie along the evenly redistributed energy
curve, while the hotter exoplanets lie nearer to the instantaneous
reradiation $T_{eq}$ curve.

Physically, $T_{eq}$ is the effective temperature
attained by an isothermal planet after it has reached complete
equilibrium with the radiation from its parent star. $T_{eq}$
is described by
\begin{equation}
\label{eq:Teq}
T_{eq} = T_{*} \left(\frac{R_{*}}{a}\right)^{1/2} [f(1-A_{\rm
B})]^{1/4} ,
\end{equation}
where $T_{*}$ and
$R_{*}$ are the effective temperature and the radius of the star, $a$
is the planet semi-major axis, and $f$ and $A_{\rm B}$ are
the re-radiation factor and the Bond albedo of the planet. 

Here we explain how hot Jupiters can exist both with and without large
day-night temperature variations. See also \cite{hube2003, harr2007,
burr2008, fort2008}.  Hot planets such as Ups And are on one side of a
temperature-driven chemical composition boundary, while cooler planets
such as HD~209458b are on the cooler side. Specifically, if the hot
Jupiter planet atmosphere is relatively cool, TiO is locked into solid
particles that have little absorbing power in the atmosphere. In the
hotter atmosphere, TiO is a ``deadly'' gas that absorbs so strongly it
puts the planet in the reradiation regime leading to a large day-night
contrast. At the temperature of these hot day side exoplanets, some
elements will be in atomic (instead of molecular) form and atomic line
opacities may also play a significant absorbing role.

What evidence do we have for the temperature-induced two-atmosphere
hypothesis? Cool stars (M stars) have
visible-wavelength spectra that are dominated---and indeed
dramatically suppressed---by TiO gas.  Hot brown dwarfs also have spectra
with TiO absorption features wheras cooler brown dwarfs do not,
implying that for cooler brown dwarfs Ti is sequestered in solid
particles. Temperature and pressures in hot Jupiter atmospheres are
similar to brown dwarfs (although the temperature gradient is
different) so that we expect a similar temperature-induced chemical
composition.

Other notable exoplanet atmosphere discoveries by {\it Spitzer} come from
spectrophotometry and include discovery of a temperature inversion on
the day side of HD~209458b \cite{burr2007, knut2008} and a tentative
detection of water vapor in transmission spectra during primary
transit \cite{tine2007,ehre2007}.

This interesting ``two types of hot Jupiter atmospheres'' hypothesis
shows just how complex hot Jupiters are. The results also imply that
3D coupled radiative transfer-atmospheric circulation models are
needed to fully understand hot Jupiters. Next generation data with
{\it JWST} (Section~\ref{sec-JWSTatm}) in terms of high SNR low-resolution
spectra as a function of orbital phase will lead to a deeper
understanding of hot Jupiter atmospheres.

\subsection{Lessons Learned from {\it Spitzer}}
\label{sec-lessonsspitzer}
{\it Spitzer} and its instruments were not designed for high contrast
observations. Here we describe some of the instrumental effects that
become important at the part-per-thousand level and consequently
affect exoplanet observations. These may be useful to consider when
planning {\it JWST} exoplanet transit observations.

The most notable instrumental effect is the ``ramp'': a gradual
detector-induced rise of up to 10\% in the signal measured in
individual pixels over time. This rise is illumination-dependent;
pixels with high levels of illumination converge to a constant value
within the first two hours of observations and lower-flux pixels
increase linearly over time.  \cite{knut2007} have attributed
this to charge trapping.  

The ramp is present at the long wavelength detectors (8 and 16 $\mu$m
and possibly also at 5~$\mu$m).  The charge trapping is likely caused by
the ionized impurities in the arsenic-doped silicon detector. The
first electrons that are released by photons get trapped by the ions
and therefore they are not immediately read out.

The ramp can be removed from a data set by a fitting a linear plus
logarithmic function.  A method to avoid the ramp is to ``preflash''
the detector before an observation, by observing a bright star. We
note that at 24~$\mu$m no ramp effect is observed \cite{demi2005};
this may be because the detector
is always illuminated by the zodiacal background. The ramp effect may
be important for {\it JWST} observations because similar detector materials
are being used on MIRI.  We recommend pre-flashing before transit
observations.

A second instrumental effect that is significant for exoplanet transit
observations is the IRAC intrapixel sensitivity variation for the
short-wavelength channels (3.5 and 4.5 $\mu$m) \cite{mora2006}.
This intrapixel sensitivity variation is a property of the detector,
is spatially asymmetric, and does not change with time. One possible
explanation of this detector effect is that there are physical gaps
between the pixels which are unresponsive to light. When an image is
centered on a pixel the highest sensitivity arises. A second
possibility is related to the bump-bond contact between the detector
and the underlying multiplexer. Each pixel makes electrical contact
with the multiplexer at pixel center.  In this scenario, the electrons
generated close to the contact point may be collected more efficiently
than electrons generated at pixel edges.  The IRAC intrapixel
sensitivity variation can be corrected for by a natural mapping of the
pixels, by exploiting the telescope pointing jitter \cite{demi2008}.

A third significant instrumental effect found from exoplanet
observations is a telescope pointing oscillation with a period of
roughly one hour. This pointing oscillation affects IRS slit spectra.
As the star becomes uncentered and recentered on the slit the number
of photons going through the slit changes. This change is
wavelength-dependent, because of the wavelength-dependent PSF,
and therefore can generate erroneous features in an
exoplanetary spectrum.  The
cause of the pointing oscillation is still a matter of debate.

A fourth effect is a longer-term telescope drift in telescope
pointing. This telescope drift is less certain than the
well-documented 1 hour oscillation. This affects the IRS spectral flux
in terms of the broad distribution of energy, because the slit width
is comparable to the diffraction width of the telescope PSF. This is
also a wavelength-dependent effect.

Despite not being designed for high-contrast observations, {\it Spitzer} has
revolutionized the study of exoplanet atmospheres by routinely
detecting secondary eclipses of hot Jupiters and measuring their
brightness temperatures at different wavelengths. Because of {\it JWST}'s
larger mirror diameter and higher spectral resolution, and by taking
care to understand and remove instrumental effects, transit
observations with {\it JWST} hold even more promise.

\section{Transiting Planet Radii with JWST}

{\it JWST} will continue to study the physical characteristics of transiting
exoplanets in the tradition of {\it Spitzer}.  See Chapter 1 of this volume
for the {\it JWST} telescope, instrument, and performance. {\it JWST} has 40 times
the collecting area of {\it Spitzer}, enabling studies of smaller
exoplanets---pushing down to potentially habitable exoplanets.

\begin{figure}[ht]
\centering
\includegraphics[scale=.45]{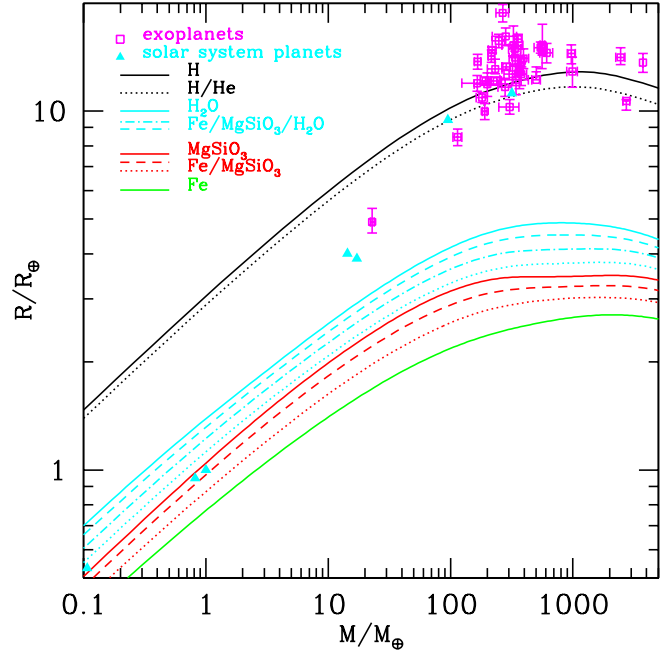}
\caption{Mass-radius relationships for solid planets.  The solid lines
are models of homogeneous planets. From top to bottom the homogeneous
planets are: hydrogen (cyan solid line); a hydrogen-helium mixture
with 25\% helium by mass (cyan dotted line); water ice (blue solid
line); silicate (MgSiO$_3$ perovskite; red solid line); and iron (Fe
($\epsilon$); green solid line). The non-solid lines are models of
differentiated planets.  The red dashed line is for silicate planets
with 32.5\% by mass iron cores and 67.5\% silicate mantles (similar to
Earth) and the red dotted line is for silicate planets with 70\% by
mass iron core and 30\% silicate mantles (similar to Mercury). The
blue dashed line is for water planets with 75\% water ice, a 22\%
silicate shell and a 3\% iron core; the blue dot-dashed line is for
water planets with 45\% water ice, a 48.5\% silicate shell and a 6.5\%
iron core (similar to Ganymede); the blue dotted line is for water
planets with 25\% water ice, a 52.5\% silicate shell and a 22.5\% iron
core. The blue triangles are solar system planets: from left to right
Mars, Venus, Earth, Uranus, Neptune, Saturn, and Jupiter. The magenta
squares denote the transiting exoplanets, including HD~149026b at 8.14
$R_{\oplus}$ and GJ~436b at 3.95 $R_{\oplus}$.  Note that electron
degeneracy pressure becomes important at high mass, causing the planet
radius to become constant and even decrease for increasing mass.  From
\cite{seag2007}.}
\label{fig:massradius} 
\end{figure}

The overall goal of planet transit observations is to combine the
radius with the planet mass to infer an interior bulk composition.
Figure~\ref{fig:massradius} shows transiting exoplanets on a
mass-radius diagram with curves for planets of homogeneous
composition. For example, we would like to know if planets in the mass
range 5 to 20 $M_{\oplus}$ have significant gas envelopes like Neptune
($\sim$ 10\% by mass), or instead consist almost
entirely of rock/iron. Owing to high temperatures at a
deeply submerged surface, the former
are not habitable, while the latter are.

\begin{figure}[ht]
\centering
\includegraphics[scale=.35]{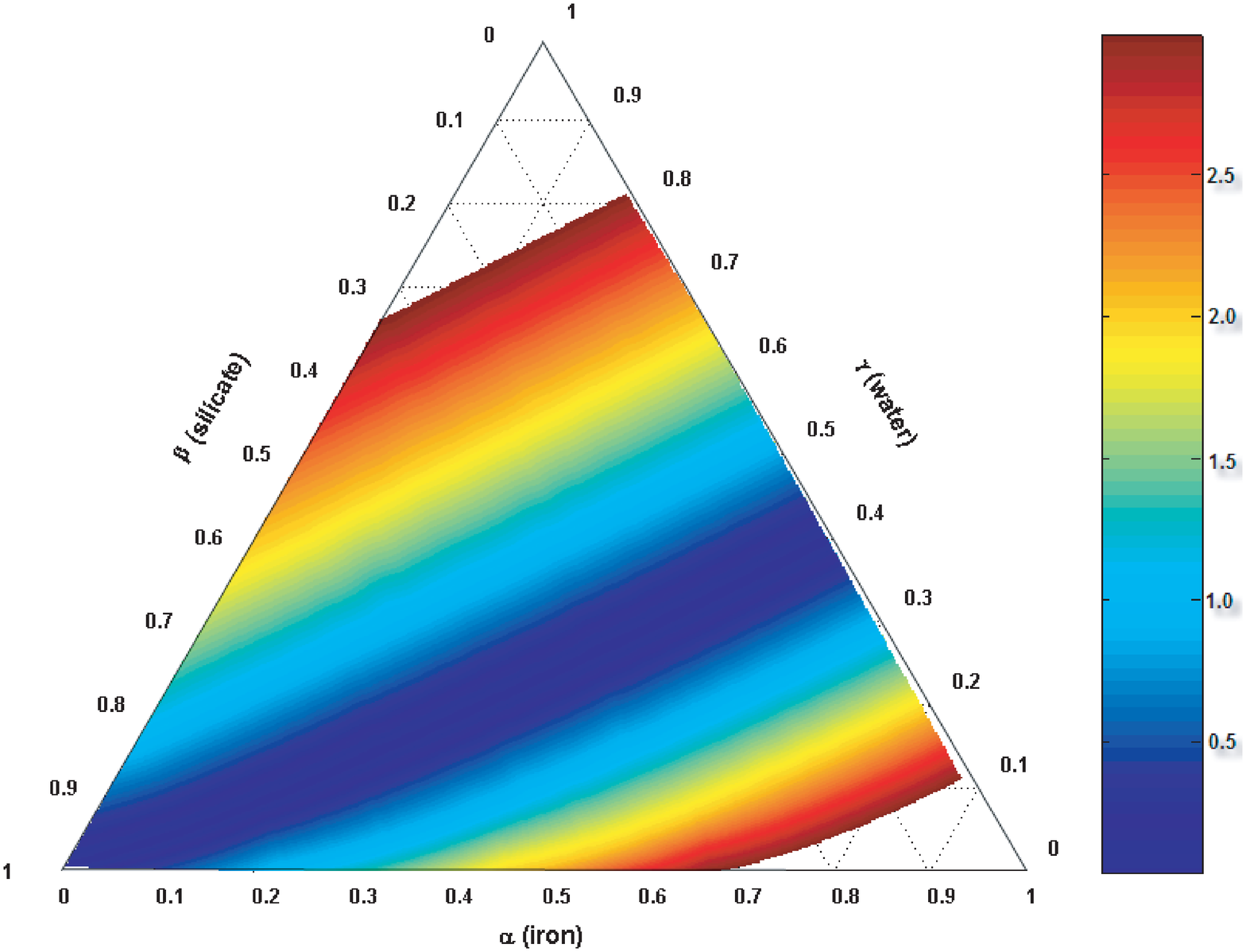}
\caption{A ternary diagram for a planet of a fixed mass and radius,
including the mass and radius uncertainties. This example is for a
planet with $M_p = 10 \pm 0.5 M_{\oplus} $ and $R_p = 2 \pm 0.1
M_{\oplus}$, showing the 1-, 2- and 3-$\sigma$ uncertainty curves as
indicated by the color bar.  Notice that considering the 3-$\sigma$
uncertainties almost the entire ternary diagram is covered---in other
words there is little constraint on the planet internal
composition. See \cite{zeng2008} for a discussion of the
direction and spacing of the curves, as well as for other examples. }
\label{fig:ternary} 
\end{figure}

To illustrate why precise radii are needed to constrain a planet's
mass we show the range of interior compositions possible for a 10
$M_{\oplus}$, 2 $R_{\oplus}$ planet on a ternary diagram
(Figure~\ref{fig:ternary}). For an explanation of ternary diagrams in
this context see \cite{vale2007} and \cite{zeng2008}. There is a
degeneracy in interior composition for a solid exoplanet made of the
three typical planetary materials: an iron core, silicate mantle, and
water outer layer. This is because of the very different densities of
the three components.  For example, a planet of a given mass and
composition could have the same radius if some of the silicate were
exchanged for a combination of water and iron. By showing contour
curves of growing observational uncertainties,
Figure~\ref{fig:ternary} emphasizes how a precise radius reduces the
interior composition uncertainty.

{\it JWST} should be able to measure precise transit light curves for a wide
range of star brightnesses. With its spectral dispersion and high
cadence observing, NIRSpec will be capable of high-precision
spectrophotometry on bright stars. NIRSpec data can then be used in
the same way that the {\it HST} STIS spectra data for HD209458 was rebinned
into ``photometry'' \cite{brow2001}.  For fainter M stars, NIRCam
should be sufficient. For all stars, the near infrared is ideal
because of negligible stellar limb darkening, removing one of the
uncertainties in deriving an accurate planet radius from a planet transit
light curve.

By scaling from the \cite{brow2001} {\it HST} data, we provide two
interesting examples of {\it JWST} precision radii \cite{jwstastrobio,
beic2007} with NIRSpec at 0.7 $\mu$m.  For the first example we
consider {\it Kepler} Earth-size planet candidates orbiting Sun-like stars
in 1 AU orbits. {\it Kepler} stars are about 300 pc distant and
Earth-analogs have a transit duration of about 8 hours.  With
high-cadence observing, {\it JWST} will be able to obtain a 35-$\sigma$
transit detection for {\it Kepler} Earth-analog planet candidates. This {\it JWST}
confirmation would be very significant, because {\it Kepler}'s SNR detection
is 7 over 4 binned transits \cite{basr2005}.

A second example is that of an Earth-sized moon orbiting the
transiting giant planet HD~209458b. At 47 pc and with a 3 hour transit
time (and 6 hour observation), {\it JWST} will also be capable of a moon
transit detection at 35-$\sigma$ SNR.

\section{Transiting Planet Atmospheres with {\it JWST}}
\label{sec-JWSTatm}
\subsection{Background}
\label{sec-background}
Transiting exoplanets present two different configurations for
atmosphere measurements. The first is during primary transit and is
called transit transmission spectra (described in
Section~\ref{sec:intro}). Transmission spectra probe the planetary
upper atmosphere and have been used to detect atomic and molecular
features in two different exoplanet atmospheres (HD~209458b and
HD~189733b), including sodium \cite{char2002}, water vapor
\cite{tine2007, swai2008}, and methane \cite{swai2008}. Additionally,
\cite{vida2003} have detected a large envelope of atomic hydrogen (and
a tentative detection of other elements) indicating a slow atmospheric
escape. \cite{redf2008} have presented the first ground-based
detection of an exoplanet atmosphere via sodium in HD~189733b.

The magnitude of the transmission spectra signal can be estimated by
the area of the planetary atmosphere compared to the area of the
star. The area of the planetary atmosphere is an annulus with a radial
height of about $5 \times H$. Here $H$ is the planetary scale height
\begin{equation}
\label{eq:H}
H = \frac{k T}{\mu_m g},
\end{equation}
where $k$ is Boltzmann's constant, $T$ is temperature, $\mu_m$ is the
mean molecular mass, and $g$ is the planet's surface gravity.
The magnitude of the planet transmission spectra signature is approximately
\begin{equation}
5 \times \frac{2 R_p H}{R_*^2},
\end{equation}
where $R_p$ is the planet's radius and $R_*$ is the star's radius. A
hot Jupiter's transmission spectra signature is approximately
10$^{-4}$. We further note that, from the planet's equilibrium 
effective temperature (equation~(\ref{eq:Teq}))
$T \sim 1/\sqrt{a}$ so that
\begin{equation}
\label{eq:transestimate}
Transmission \sim \frac{1}{\sqrt{a}}.
\end{equation}

We emphasize a very critical difference between planets with
hydrogen-rich atmospheres (including both the upper layers of giant
planet envelopes and thinner atmospheres of super Earths) and
terrestrial planets (including Earths or super Earths) with relatively
thin N$_2$ or CO$_2$ atmospheres. The factor $\mu_m$ is 2 for an H$_2$
atmosphere, but 44 for a CO$_2$ atmosphere!  Hence, the difference in
transmission spectra between hydrogen-rich atmospheres and
terrestrial-like planet atmospheres is a factor of 20 (with $T$ and
$g$ being equal; see \cite{mill2008} for further discussion). This
implies that, while smaller space telescopes can study hydrogen-rich
exoplanet atmospheres, {\it JWST}'s 6.5 m effective mirror diameter is
needed to study CO$_2$- or N$_2$-dominated atmospheres similar to
terrestrial planet atmospheres in our solar system.

The second configuration available for transiting atmosphere studies
with {\it JWST} is secondary eclipse. Observations during secondary
eclipse measure the planet's thermal emission. This contains
information about the planet's temperature and temperature
gradient. Spectral features can also be detected with the planet's
thermal emission flux. If absorption lines are detected, the planet
has a temperature that is decreasing towards the top of the
atmosphere. If emission lines are detected, the planet has a
temperature that is increasing towards the top of the atmosphere.

Estimating the magnitude of the planet's thermal emission (in the form
of a planet-to-star flux ratio) is not easy because planet model
atmospheres are usually needed. That said, we can bracket an estimate
with two extremes. One extreme is the case where the thermal emission
spectra could be observed over a broad infrared wavelength range to
estimate the ``bolometric'' planet flux.  In this case, we can
estimate the planet-star flux ratios by a ratio of black bodies, and
considering the Stefan-Boltzmann law $F = \sigma_R T^4$,
\begin{equation}
Emission \sim \frac{R_p^2 T_p^4}{R_*^2 T_*^4},
\end{equation}
where we have written $T_p = T_{eq}$.
Again using the scaling relation
$T_p \sim 1/\sqrt{a}$, we find 
\begin{equation}
\label{eq:emis1}
Emission \sim \frac{1}{a^2}.
\end{equation}
As a separate extreme to estimate the thermal
emission planet-star contrast ratio, we can take the Rayleigh-Jeans tail
of the black body spectrum $h \nu \ll kT$ to get
\begin{equation}
Emission = \frac{R_p^2 T_p}{R_*^2 T_*},
\end{equation}
and again using the $T_{eq} \sim 1/\sqrt{a}$ scaling,
\begin{equation}
\label{eq:emis2}
Emission \sim \frac{1}{\sqrt{a}}.
\end{equation}
More reasonably, we can assume that at the peak of the planet's
output, which is neither represented by the bolometric flux nor is it
in the Rayleigh-Jeans tail (see Figure~\ref{fig:blackbody}), the
dependence with $a$ falls between that of $\sim 1/a^2$
(equation~(\ref{eq:emis1})) and $\sim 1/\sqrt{a}$
(equation~(\ref{eq:emis2})).

We have gone through these estimates to make a single main point: a
comparison between the semi-major axis ($a$) dependence of
transmission and emission spectra.  Emission spectra have a stronger
signal than transmission spectra for planets orbiting close to their
parent stars (and indeed emission spectra are only possible for
planets close to their stars). While transmission spectra are weaker
than emission spectra for planets close to their stars, they are still
attainable for planets orbiting far from their stars.

\subsection{Giant Planet Atmospheres}

The {\it JWST} thermal IR detection capability can be explored by
scaling the {\it Spitzer} results (this discussion is from
\cite{beic2007}). The 5-8~$\mu$m range is ideal for solar-type stars
because the planet-star contrast is high and the exo-zodiacal
background is low. For an estimate we can take the TrES-1 5$\sigma$
detection at 4.5 $\mu$m, taking into account that {\it JWST} has 40
times the collecting area of {\it Spitzer} and assuming that the
overall efficiency of {\it JWST} is almost 2x lower, giving an
effective collecting area improvement of $\sim$25 times. {\it JWST}
will therefore be able to detect hot Jupiter thermal emission at an
SNR of 25 around stars at TrES-1's distance ($\sim$150 pc; a distance
that includes most stars from shallow ground-based transit
surveys). Similarly, {\it JWST} can detect a hot planet 5 times
smaller than TrES-2, or down to 2 Earth radii, for the same set of
stars, assuming that instrument systematics are not a limiting
factor. Scaling with distance, {\it JWST} can detect hot Jupiters
around stars 5 times more distant than TrES-1 to SNR of 5, which
includes all of the {\it Kepler} and {\it COROT} target stars. Beyond
photometry, {\it JWST} can obtain thermal emission spectra (albeit at
a lower SNR than for photometry of the same planet). Rebinning the
R=3,000 NIRSpec data to low-resolution spectra will enable detection
of H$_2$O, CO, CH$_4$, and CO$_2$.

For hot Jupiter transmission spectra, we turn to simulations of
NIRSpec transmission spectra \cite{vale2005}.  NIRSpec will have three
``high-resolution'' (2200 $<$ R $<$ 4400) gratings (G140H, G235H,
G395H) that span the wavelength range 1-5~$\mu$m. Observations of
planet host stars will be made with the largest fixed aperture
(4''$\times$0.4''), rather than microshutters. A 2048$\times$64
subarray on each of the two detectors will be nondestructively read
``up-the-ramp'' every 0.3 s for 2.4 s (G140H, G235H) or 3.6 s
(G395H). The detector is then reset to avoid charge saturation ($>$
60,000 e$^-$), and the process is repeated thousands of times to build
up 2 hours of total exposure time. The total spacecraft time required
to achieve this exposure time depends on the idle time between
individual subarray exposures. {\it JWST} has the potential to characterize
the atmospheres of dozens of transiting planets, if overheads and
calibration errors can be controlled.

One of us \cite{vale2005} has simulated NIRSpec observations with {\it
JWST}. They estimated NIRSpec performance based on observatory
requirements and some lab data. For reference, {\it JWST} has 25~m$^2$
clear aperture and a 33\% peak efficiency for the NIRSpec/G140H mode
considered here. The simulated observations include those in and out
of transit, based on NextGen models \cite{haus1999} and planetary
absorption spectra from \cite{brow2001b}. Simulated noise takes into
account the details of how the detector will be read and how spectra
will be extracted. (This software is available upon request from
valenti@stsci.edu.)  \cite{vale2005} do not attempt to estimate the
impact of systematic errors that will undoubtedly affect actual
observations.

\begin{figure}[ht]
\centering
\includegraphics[scale=.5]{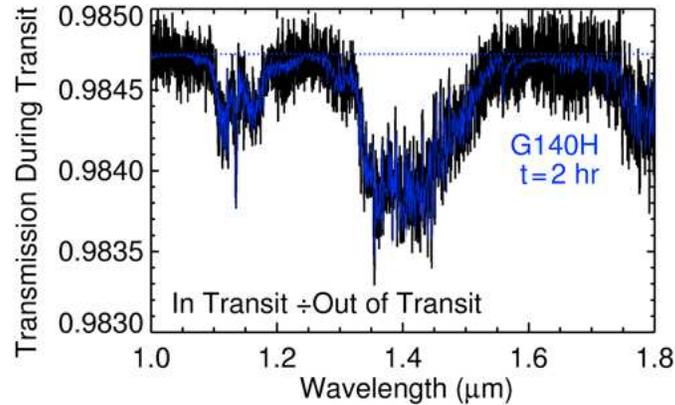}
\caption{Simulated NIRSpec measurements of transmission spectra of a
hot Jupiter exoplanet.  The spectrum is the ratio of the simulated
spectrum in and out of transit.  The blue spectrum shows the Brown et
al. (2001) model used to generate the simulated observations (black
curve).  The dominant absorption features are water vapor.  }
\label{fig:HJspectra} 
\end{figure}

Figure~\ref{fig:HJspectra} shows a simulation of a $K=12$, $V$=$13.4$
star with a 6 hour observation using NIRSpec/G140H, centered on a 2h
planetary transit. Both the theoretical spectra and the simulated
observations are shown. In the simulated observation, the water vapor
absorption features are obvious and their detection significance is
high. This is in contrast to the 3 to 4 SNR transmission spectra
detections measured with {\it HST}, {\it Spitzer}, and from the ground.

A magnitude of $V$=$13.4$ encompasses most stars surveyed in the shallow
ground-based transit surveys, which may discover well over 100 hot Jupiters
by the time of {\it JWST}'s launch. {\it Kepler}'s star magnitude range is 
$V$=$9$ to 16 (most transits will be found around stars $V$=14 to 16), and
therefore the simulated spectrum in Figure~\ref{fig:HJspectra} applies
to {\it Kepler}'s hot Jupiters as well (although with a slightly lower SNR
for the fainter-end stars).  By binning NIRspec data in wavelength,
Jupiter transiting planets orbiting out to 1 AU are accessible for
NIRSpec transmission spectra.  Such observations may help resolve the
puzzling question on the origin of the ``puffed-up'' hot Jupiters that
are too large for their mass and age (upper right corner in
Figure~\ref{fig:massradius}).

Other outstanding questions for hot Jupiters that {\it JWST} can address
include the atmospheric circulation. Tidally-locked to their parent
stars, hot Jupiters have a permanent day and night side. By studying
the thermal emission spectra as a function of phase we can get a
handle on how the temperature of different layers of the planet is
changing. Intriguing are the eccentric hot Jupiters whose atmospheres
are intensely heated for a brief period of time. Spectra as a function of
phase will help us to determine the radiation time constant, a 
fundamental factor in understanding atmospheric circulation.

By binning NIRSpec data in wavelength, transiting planets smaller than
Jupiter can be studied as well. Between Jupiters and super Earths, we
expect many hot transiting Neptune-sized exoplanets to be known and
accessible to study by {\it JWST}.

\subsection{Terrestrial Planet Atmospheres}
\label{sec-habitable}

One of the most interesting exoplanet questions {\it JWST} can address
is whether a planet is habitable. By habitable, we mean, in the
conventional sense, one with surface liquid water. Atmospheric water
vapor is a good indication of surface liquid water. On Earth, O$_2$ is
considered the most significant biosignature, given that it is a
highly reactive gas with a very short lifetime, and only produced in
large quantities by biological processes.  For signs of life, one
would ideally want to observe signatures of molecules that are highly
out of redox equilibrium (such as methane and oxygen)\cite{lede1965,
love1965}, but in reality it will be difficult enough to observe any
molecular signature robustly from an Earth-temperature, near
Earth-size exoplanet. Finally, CO$_2$, while not a biosignature,
indicates an terrestrial-planet atmosphere.  See \cite{desm2002} for a
discussion of Earth's biosignatures.

We choose six examples to illustrate {\it JWST}'s capabilities for studying
terrestrial exoplanets orbiting in their star's habitable zone.

The first is for a large Earth-like exoplanet. In \cite{jwstastrobio},
Gilliland considered a 1.5 Earth-radius planet orbiting a Sun-like
star at 1 AU at 20 pc distance.  {\it JWST} could achieve a 500-$\sigma$
detection with 0.25 sec exposures, 0.5 sec down time, and $3 \times
10^8$ photons per integration (i.e., $10^5$ pixels for 50k electrons
in the brightest pixel). This kind of detection can distinguish
between Earth- and Venus-like atmospheres. The observations required
for such a 4-$\sigma$ discrimination in this case would span 30 hours
centered on a 10-hour transit.  This measurement relies on a {\it JWST}
capability of efficiently recording high photon fluxes. More
significantly, this hypothetical planet-star system is optimistic;
transiting Earths around $V$=6 dwarfs will be rare and currently no
transit survey is capable of finding them.

The second and third examples are for Earth-size planets orbiting M
stars. There is new excitement in finding transiting planets orbiting
in the habitable zones of M stars. Recall that the habitable zone is
defined as the location around a star where a planet's surface
temperature will permit liquid water.  Owing to their low luminosity,
M stars have habitable zones much closer (3 to 40 day period orbits)
than Sun-like stars (1 year orbits).  In comparison to the above
example of a 1.5 Earth-radius planet in a 1 AU orbit transiting a
Sun-like star: the probability to transit is high (transit probability
is $R_*/a$); transits are deep; the radial velocity signature is
higher and mass measurements are possible; the large planet-star
contrast may permit thermal emission measurements.  A
ground-based targeted transit search for bright M stars is underway
\cite{nutz2007} and expects to yield a few potentially habitable
planets, as are ongoing radial velocity surveys of M stars.

\begin{figure}[ht]
\includegraphics[scale=.4]{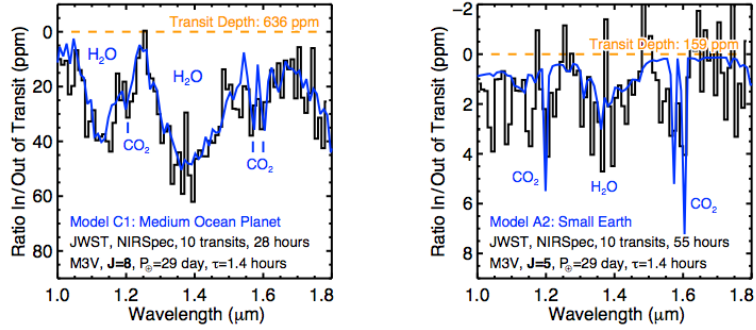}
\caption{Simulated NIRSpec measurements of transmission spectra of
habitable exoplanets. The spectra are the ratio of simulated spectra
in and out of transit. Details of the planet and star scenarios are
indicated on the Figure. The blue spectrum shows the Ehrenreich et
al. (2006) models used to generate the simulated observations (black
curve). The dominant absorption features are indicated.}
\label{fig:EarthNIRSpec} 
\end{figure}

One of us \cite{valenti2007} has simulated NIRSpec spectra for
transiting Earth-like spectra \cite{ehre2006} orbiting bright M stars
(Figure~\ref{fig:EarthNIRSpec}). Orbiting in the habitable zone (P=29
days) of an M3V, J=8 star, an ocean planet will have a strong
atmospheric water vapor signature at the level of 40 to 60 ppm. CO$_2$
features may also be detectable.  With a 1.4 hour transit duration, 10
transits and 28 hours of observing time are needed for suitable SNR.
This ocean planet example is an 0.5 $M_{\oplus}$, 1 $R_{\oplus}$
planet.  Its lower density makes its scale height more than 2 times
higher than Earth's (Section~\ref{sec-background}) making a
transmission spectrum twice as easy to detect (see
Section~\ref{sec-background}). For a ``small Earth'' (defined as 0.1
Earth masses and 0.5 Earth radii) also orbiting in the habitable zone
of M3V star (but now for a slightly brighter, J=6 star), the water
vapor signatures are several ppm. 54 hours of observing time would be
needed for a significant detection. With the period of 29 days in both
of these examples, scheduling to observe 10 or more transits is
critical.  We emphasize that more work both in modeling the planet
atmospheres and in {\it JWST} simulations needs to be done. Regardless, M3V
or later stars as bright as J=6 to J=8 are rare, making transiting
planets even more rare. 

The fourth example is for a super Earth with a hydrogen-rich
atmosphere instead of a hydrogen-poor atmosphere. A hydrogen-rich
atmosphere would be created by outgassing on a super Earth with a
surface gravity high enough to prevent loss of all of the hydrogen. We
have described in Section~\ref{sec-background} how the scale height is
inversely proportional to surface gravity and mean molecular
weight. We take a 5~$M_{\oplus}$, 1.5 $R_{\oplus}$ planet
\cite{seag2007} with a corresponding surface gravity 2.2 times higher
than Earth's and a mean molecular weight 44 times lower.  The
transmission spectra signal will be 10 times stronger than the medium
ocean planet described in example 2 and
Figure~\ref{fig:EarthNIRSpec}. Such a hydrogen-rich kind of planet
would require $\sim$ three times less observing time than the
Earth-like planet atmosphere in example 2 above, or approximately 10
hours. For further discussion on hydrogen-rich atmospheres on super
Earths see \cite{mill2008}.

\begin{figure}[ht]
\centering
\includegraphics[scale=0.5]{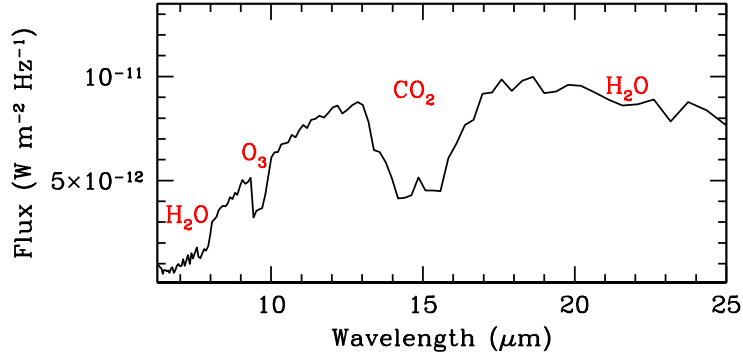}
\caption{Earth's mid-infrared spectrum as observed by Mars Global
Surveyor enroute to Mars \cite{chri1997}. Major molecular absorption features
are noted.}
\label{fig:Earth} 
\end{figure}

For our final examples, we consider the possibility of mid-IR thermal
emission detection during secondary eclipse of an Earth-temperature
planet with MIRI. MIRI spectra, in general, may be very useful because
with no slit, no detrimental effect from pointing errors will occur
(see Section~\ref{sec-lessonsspitzer}).  Regarding Earth-type planets,
detection of the ozone (O$_3$) and CO$_2$ spectral features are key
(Figure~\ref{fig:Earth}).

One of us (Deming) has simulated MIRI observations (also presented in
\cite{char2007b} but with minor corrections here). The planet is
modeled as a black body in thermal equilibrium with its star, using a
Kurucz model atmosphere for the star \cite{kuru1992}.  Since the star
is bright, its statistical photon fluctuations are a dominant noise
source.  Noise from the thermal background of the telescope and Sun
shield, and background noise from zodiacal emission in our solar
system are also included.  The efficiency of the telescope/MIRI
optical system (electrons out / photons in) was taken to be 0.3.

\begin{figure}[ht]
\centering
\includegraphics[scale=0.4]{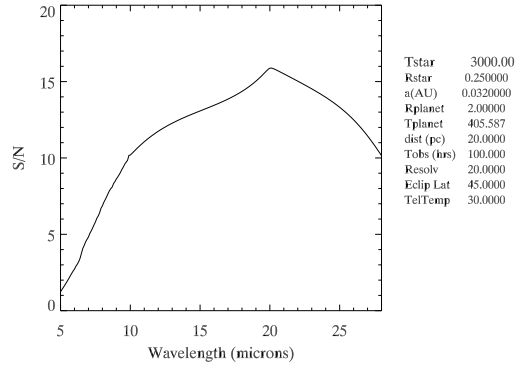}
\vspace{0.4in}
\caption{MIRI simulation of the SNR as a function of
wavelength for R=20 spectroscopy of a 2 $R_{\oplus}$ super Earth
orbiting at 0.03 AU around an 20-pc-distant M5V host star (where the
planet's $T_{eq}$ = 406~K).  The SNR would be high enough to detect
the O$_3$ or CO$_2$ features shown in Figure~\ref{fig:Earth}.}
\label{fig:miri1} 
\end{figure}

\begin{figure}[ht]
\centering
\includegraphics[scale=0.4]{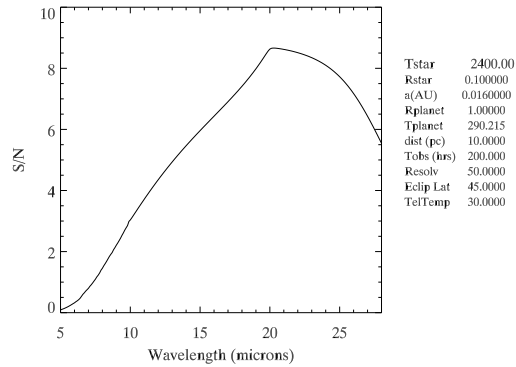}
\vspace{0.4in}
\caption{MIRI simulation of the SNR as a function of wavelength for
R=50 spectroscopy of a 1 $R_{\oplus}$ planet orbiting at 0.016 AU
around an 10-pc-distant M8V host star (where the planet's $T_{eq}$ =
290~K).  The SNR would be high enough to detect the O$_3$ or CO$_2$
features shown in Figure~\ref{fig:Earth}.}
\label{fig:miri2} 
\end{figure}

Figure~\ref{fig:miri1} shows the SNR for MIRI R=20 spectroscopy of a 2
$R_{\oplus}$ super Earth orbiting at 0.03 AU around a 20-pc-distant
M5V host star as a function of wavelength. Figure~\ref{fig:miri1} aims to show
that the SNR would be high enough to detect the O$_3$ or CO$_2$
features shown in Figure~\ref{fig:Earth}. With a 100 hour observation
(for $\sim$40 transits and with the total time divided between
in-eclipse and out-of eclipse) a SNR of 10 to 15 is possible.
Figure~\ref{fig:miri2} shows a similar example, but for a 200 hour
observation of a 1 $R_{\oplus}$ planet orbiting in the habitable zone
of a 10-pc-distant M8V star with R=50.  For comparison, we note that
100 hours is a bit less than half of the Hubble Deep Field observing
time.

For both SNR vs. wavelength examples in Figures~\ref{fig:miri1} and
\ref{fig:miri2}, the SNR increases at the blue wavelength range due to
the increasing number of photons from the planet. At longer
wavelengths, the SNR decreases because of greater thermal background
noise from the telescope and the Sun shield. Like the above 
transmission spectra examples, transiting exoplanets
suitable for {\it JWST}/MIRI followup observations of secondary eclipses
are anticipated to be rare.

The only {\it JWST} instrument we have not discussed
is NIRCam, simply due to the current lack of available exoplanet
simulations.  A brief discussion of transiting exoplanet science with
the NIRCam grisms is presented in \cite{gree2008}. NIRCam will be
suitable for both observing primary transit and secondary eclipse. Due
to the lack of slit (see Section~\ref{sec-lessonsspitzer}), the
grisms, and the near-infrared wavelengths where many molecules have
absorption features, NIRCam holds huge promise for transiting
exoplanet atmosphere studies.

In summary, {\it JWST} has the capability to study spectral features of
Earth-size or larger planets in the habitable zones of main sequence
stars. We need to first find these rare transiting planets and second
be patient with the tens to hundreds of hours of {\it JWST} time with the
concomitant complex scheduling to cover periodic transits.

\section{Discussion}

The {\it Hubble Space Telescope} and the {\it Spitzer Space Telescope}
(Section~\ref{sec-lessonsspitzer}) were not designed to achieve the
very high SNRs necessary to study transiting exoplanets, but they have
succeeded nonetheless.  We have learned from these observations that
with enough photons, systematics that were unknown in advance can
often be corrected for (as long as the errors are uncorrelated).
Given the expected thermal stability of {\it JWST} and the differential
nature of transit observations, we are optimistic that {\it JWST}, too, will
succeed in high-contrast transit observations.

We have described a few examples where {\it JWST} will have a significant
impact, including measuring precise exoplanet radii for all sizes of
exoplanets and spectra for giant planets at a variety of semi-major
axes. Individually, these observations will be ``cheap'' in terms of
telescope time, with single transit measurements being sufficient.

The most significant exoplanet observations {\it JWST} is poised to make are
those for potentially habitable exoplanets. In the most optimistic
case, {\it JWST} is able to identify planets with atmospheric water vapor,
or even chemical disequilibrium indicative of biological origin. {\it JWST}
has the capability to do so if the tens to hundreds of hours per
target are allocated, and if such rare transiting planets can be
discovered in sufficient numbers.

\begin{acknowledgement}
We thank Mark Clampin, George Ricki, Dave Charbonneau, and Heather
Knutson for useful discussions.

\end{acknowledgement}


\begin{thebibliography}{99}

\bibitem[1]{exoe} http://exoplanet.eu/

\bibitem[2]{butl2006} Butler, R. P. et al. 2006, ApJ, 646, 505

\bibitem[3]{udry2007} Udry, S., Fischer, D., 
\& Queloz, D.\ 2007, Protostars and Planets V, 685 

\bibitem[4]{char2007a} Charbonneau, D., 
Brown, T.~M., Burrows, A., 
\& Laughlin, G.\ 2007, Protostars and Planets V, 701 

\bibitem[5]{gaud2008} Gaudi, B.~S., et al.\ 
2008, Science, 319, 927 

\bibitem[6]{chau2004} Chauvin, G., Lagrange, 
A.-M., Dumas, C., Zuckerman, B., Mouillet, D., Song, I., Beuzit, J.-L., \& 
Lowrance, P.\ 2004, A\&A, 425, L29 

\bibitem[7]{wols1992} Wolszczan, A., \& Frail, D.~A.\ 1992, Nature, 355, 145 

\bibitem[8]{silv2007} Silvotti, R., et al.\ 2007, Nature, 449, 189

\bibitem[9]{chau2005} Chauvin, G., et al.\ 2005, A\&A, 438, L29 

\bibitem[10]{neuh2005} Neuh{\"a}user, 
R., Guenther, E.~W., Wuchterl, G., Mugrauer, M., Bedalov, A., \& 
Hauschildt, P.~H.\ 2005, A\&A, 435, L13 

\bibitem[11]{gard2006} Gardner, J.~P., et al.\ 
2006, Space Science Reviews, 123, 485 

\bibitem[12]{irac} Fazio, G.~G., et al.\ 
2004, ApJS, 154, 10 

\bibitem[13]{irs} Houck, J.~R., et al.\ 
2004, ApJS, 154, 18 

\bibitem[14]{mips} Rieke, G.~H., et al.\ 
2004, ApJS, 154, 25 

\bibitem[15]{rich2006} Richardson, L.~J., 
Harrington, J., Seager, S., \& Deming, D.\ 2006, ApJ, 649, 1043 

\bibitem[16]{butl2004} Butler, R.~P., Vogt, 
S.~S., Marcy, G.~W., Fischer, D.~A., Wright, J.~T., Henry, G.~W., Laughlin, 
G., \& Lissauer, J.~J. 2004, ApJ, 617, 580 

\bibitem[17]{gill2007} Gillon, M., et al. 2007, A\&A,
472, L13

\bibitem[18]{demi2007} Deming, D., Harrington, 
J., Laughlin, G., Seager, S., Navarro, S.~B., Bowman, W.~C., \& Horning, 
K.\ 2007, ApJ, 667, L199 

\bibitem[19]{gill2007b} Gillon, M., et al.\ 
2007, A\&A, 471, L51 

\bibitem[20]{cox2000} Cox, A.~N.\ 2000, Allen's 
astrophysical quantities, 4th ed.~Publisher: New York: AIP Press; Springer, 
2000.~Editedy by Arthur N.~Cox.~ ISBN: 0387987460  

\bibitem[21]{knut2007} Knutson, H.~A., et al.\ 
2007, Nature, 447, 183 

\bibitem[22]{demi2005} Deming, D., Seager, S., 
Richardson, L.~J., \& Harrington, J.\ 2005, Nature, 434, 740 

\bibitem[23]{harr2006} Harrington, J., 
Hansen, B.~M., Luszcz, S.~H., Seager, S., Deming, D., Menou, K., Cho, 
J.~Y.-K., \& Richardson, L.~J.\ 2006, Science, 314, 623 

\bibitem[24]{cowa2007} Cowan, N.~B., Agol, E., 
\& Charbonneau, D.\ 2007, MNRAS, 379, 641 

\bibitem[25]{seag2005} Seager, S., Richardson, L.~J., Hansen,
B.~M.~S., Menou, K., Cho, J.~Y.-K., \& Deming, D.\ 2005, ApJ, 632,
1122

\bibitem[26]{harr2007} Harrington, J., Luszcz, S., Seager, S., Deming,
D., \& Richardson, L.~J.\ 2007, Nature, 447, 691

\bibitem[27]{burr2008} Burrows, A., Budaj, J., \& Hubeny, I. 2008,
ApJ, 678, 1436

\bibitem[28]{fort2008} Fortney, J.~J., Lodders, K., Marley, M.~S., \&
Freedman, R.~S.\ 2007, ArXiv e-prints, 710, arXiv:0710.2558

\bibitem[29]{show2007} Showman, A.~P., Menou, 
K., \& Y-K.~Cho, J.\ 2007, ArXiv e-prints, 710, arXiv:0710.2930 

\bibitem[30]{hube2003} Hubeny, I., Burrows, A., \& Sudarsky, D.\ 2003, ApJ, 594, 1011

\bibitem[31]{char2005} Charbonneau, D., et al.\ 2005, ApJ, 626, 523

\bibitem[32]{burr2007} Burrows, A., Hubeny, I., Budaj, J., Knutson,
H.~A., \& Charbonneau, D.\ 2007, ApJ, 668, L171

\bibitem[33]{knut2008} Knutson, H.~A., 
Charbonneau, D., Allen, L.~E., Burrows, A., \& Megeath, S.~T.\ 2008,
ApJ, 673, 526 

\bibitem[34]{tine2007} Tinetti, G., et al.\ 
2007, Nature, 448, 169 

\bibitem[35]{ehre2007} Ehrenreich, D., 
H{\'e}brard, G., Lecavelier des Etangs, A., Sing, D.~K., D{\'e}sert, J.-M., 
Bouchy, F., Ferlet, R., \& Vidal-Madjar, A.\ 2007, 
ApJ, 668, L179 

\bibitem[36]{mora2006} 
Morales-Calder{\'o}n, M., et al.\ 2006, ApJ, 653, 1454 

\bibitem[37]{demi2008} 
Deming, D., \& Seager, S., submitted to ApJ

\bibitem[38]{seag2007} Seager, S., Kuchner, M., 
Hier-Majumder, C.~A., \& Militzer, B.\ 2007, ApJ, 669, 1279 

\bibitem[39]{vale2007} Valencia, D., 
Sasselov, D.~D., \& O'Connell, R.~J.\ 2007, ApJ, 665, 1413 

\bibitem[40]{zeng2008} Zeng and Seager, PASP, in press

\bibitem[41]{brow2001} Brown, T.~M., 
Charbonneau, D., Gilliland, R.~L., Noyes, R.~W., \& Burrows, A.\ 2001, 
ApJ, 552, 699 

\bibitem[42]{jwstastrobio} Seager and Lunine, eds. 
{\it JWST} and Astrobiology white paper, NASA Astrobiology Institute

\bibitem[43]{beic2007} Beichman, C.~A., 
Fridlund, M., Traub, W.~A., Stapelfeldt, K.~R., Quirrenbach, A., \& Seager, 
S.\ 2007, Protostars and Planets V, 915 

\bibitem[44]{basr2005} Basri, G., Borucki, 
W.~J., \& Koch, D.\ 2005, New Astronomy Review, 49, 478 

\bibitem[45]{char2002} Charbonneau, D., 
Brown, T.~M., Noyes, R.~W., \& Gilliland, R.~L.\ 2002, ApJ, 568, 377 

\bibitem[46]{swai2008} Swain, M. R., Vasisht, G., \& Tinetti, G.\ 2008, 
Nature, 452, 329

\bibitem[47]{vida2003} Vidal-Madjar, A., 
Lecavelier des Etangs, A., D{\'e}sert, J.-M., Ballester, G.~E., Ferlet, R., 
H{\'e}brard, G., \& Mayor, M.\ 2003, Nature, 422, 143 


\bibitem[48]{redf2008} Redfield, S., Endl, 
M., Cochran, W.~D., \& Koesterke, L.\ 2008, ApJ, 673, L87 

\bibitem[49]{mill2008} Miller-Ricci, Seager, Sasselov,
submitted to ApJ

\bibitem[50]{vale2005} Valenti, J.~A., et al.\ 
2005, Bulletin of the American Astronomical Society, 37, 1350 


\bibitem[51]{haus1999} Hauschildt, P.~H., 
Allard, F., \& Baron, E.\ 1999, ApJ, 512, 377 

\bibitem[52]{brow2001b} Brown, T.~M.\ 2001, ApJ, 553, 
1006 

\bibitem[53]{lede1965} Lederberg, J. \ 1965,
 Nature, 207, 9 

\bibitem[54]{love1965} Lovelock, J.~E. \ 1965,
 Nature, 207, 568 

\bibitem[55]{desm2002} Des Marais, D.~J., 
et al.\ 2002, Astrobiology, 2, 153 

\bibitem[56]{nutz2007} Nutzman, P., \& 
Charbonneau, D.\ 2007, ArXiv e-prints, 709, arXiv:0709.2879 

\bibitem[57]{valenti2007} Valenti, J.~A., 
Turbull, M., McCullough, P., \& Gilliland, R. \ 2008,
in prep.

\bibitem[58]{ehre2006} Ehrenreich, D., 
Tinetti, G., Lecavelier Des Etangs, A., Vidal-Madjar, A., \& Selsis, F.\ 
2006, A\&A, 448, 379 

\bibitem[59]{chri1997} Christensen, 
P.~R., \& Pearl, J.~C.\ 1997, JGR, 102, 10875 

\bibitem[60]{char2007b} Charbonneau, D., 
\& Deming, D.\ 2007, ArXiv e-prints, 706, arXiv:0706.1047 

\bibitem[61]{kuru1992} Kurucz, R.~L.\ 1992, The 
Stellar Populations of Galaxies, 149, 225 

\bibitem[62]{gree2008} Greene, T., et al. \ 2008, SPIE, in press


\end{thebibliography}
\end{document}